\documentclass[reprint,amsmath,amssymb,aps]{revtex4-1}

\usepackage{graphicx}
\usepackage{color}
\usepackage{dcolumn}
\usepackage{bm}
\usepackage[mathlines]{lineno}
\newcommand{\B}{\mathcal{B}}
\newcommand{\E}{\mathcal{E}}
\newcommand{\hil}{\mathcal{H}}
\newcommand{\ketbra}[1][]{| #1 \rangle \langle #1 |}
\newcommand{\ket}[1][]{| #1 \rangle}
\newcommand{\bra}[1][]{\langle #1 |}
\newcommand{\vect}[1][]{\mathbf{ #1 }}
\newcommand{\vecn}{{\mathbf{n}}}

\newcommand{\trace}{\mathrm{Tr}}
\newcommand{\id}{\mathbb{I}}
\newcommand{\dif}{\mathrm{d}}
\newcommand{\rhn}{\rho_{\mathbf{n}}}
\newcommand{\rhmn}{\rho_{- \mathbf{n}}}

\newcommand{\hfe}{\tau_{\mathbf n, \delta}}
\newcommand{\fidelity}{{\mathrm F}}
\newcommand{\pauliy}{\sigma_2}

\newcommand{\sphere}{{{\mathbf n} \in S_2}}
\newcommand{\ensrhn}{{\mathcal{E}_\rho}}
\newcommand{\enshfe}{{\mathcal{E}_\tau}}
\newcommand{\ssm}{{\mathcal{M}}}
\newcommand{\ptp}{{\mathcal{L}}}
\newcommand{\rhne}{\rho_{\mathbf{n},\delta}}
\newcommand{\rhmne}{\rho_{- \mathbf{n},\delta}}
\newcommand{\sr}{2}
\newcommand{\dep}{\mathcal{D}}
\newcommand{\para}{{\psi^{\mathrm{para}}}}
\newcommand{\anpa}{{\psi^{\mathrm{anti}}}}

\begin{document}
\title{Strongly Non-Quantitative Classical Information in Quantum Carriers}
\author{Jisho Miyazaki}
\affiliation{Oi-cho, Fukui, 919-2125, Japan}
\date{\today}

\begin{abstract}
A simple method to enhance the quality of communication is to send a carrier with its copies.
Classical information theory says that information behaves quantitatively under copying.
In other words, if a carrier is more informatic than another carrier, it remains so when they are compared with their copies.
Using the lens of quantum mechanics, we challenge this accepted fact of classical information theory.
Specifically, we examine two quantum systems parameterized differently by the same random variable such that the first system alone offers a more accurate guess about the variable in any figure of merit, while the two copies of the second system together do more in some figures of merit than the two copies of the original system.
This finding unveils a conceptual discrepancy between classical information and its carrier, and implies the possibility of hiding classical information in a form of quantum information.
\end{abstract}

\maketitle

\section{introduction}
When a complete description for a carrier of information is given by a probability distribution or a quantum density operator, a single measurement of the carrier may not be sufficient to perfectly recover original information conveyed by the carrier \cite{Helstrom1976,Ivanovic1987,Dieks1988,Peres1988,Crokeetal2006}.
It is better to request multiple copies of the same carrier from the source if possible.
If we have limitations on resources such as the number of copies, we have to optimize the measurement and guessing strategy for better information.
The observer can make increasingly accurate guesses about the original information when optimized measurements are performed on an increasing number of copies.

In this article, we show that the amount of information contained in quantum carriers may increase under copying so that it behaves non-quantitatively with respect to the number of copies to be measured together.
Suppose we have two carriers $\ensrhn$ and $\enshfe$, whose states are differently parameterized by a random variable of the underlying physical system.
Carrier $\ensrhn$ alone is assumed to offer better knowledge about the system than carrier $\enshfe$ does: that is, the reader can make a more accurate guess about the value of random variables from measurement results on $\ensrhn$ than from on $\enshfe$, where the accuracy is measured by a certain figure of merit.
The reader might guess that multiple copies of $\ensrhn$ will give even better information than multiple copies of $\enshfe$, and would prefer to have carrier $\ensrhn$ no matter whether copying is possible or not.
Behind this guess is an intuition that the information content is a quantity inherent to its carriers, and grows quantitatively (though not proportionally) along the number of identical carriers.

\begin{figure}[htbp]
	\centering
	\includegraphics[width=8.6cm]{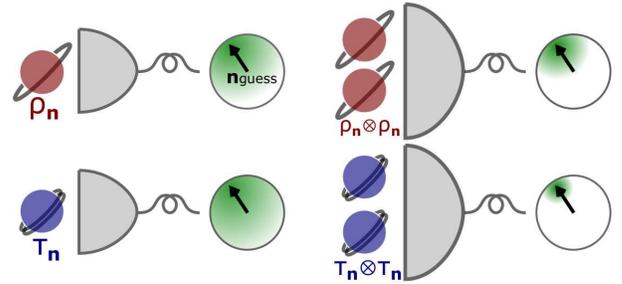}
	\caption{\label{fig:carriers} A schematic presentation of non-quantitative information.}
\end{figure}

However, if the carriers are quantum entities, two copies of $\enshfe$ may offer better knowledge about the system as depicted in FIG.~\ref{fig:carriers}.
A series of analyses on entangled measurements \cite{PeresWootters1991,Massar1995,Massar2000,Prydeetal2005,Nisetetal2007} leads to the existence of two carriers with the following property.
The first alone contains more information in a certain measure.
The two copies of second carrier get the benefit of entangled measurement and together offer more information in the same measure than the two copies of the first carrier.
When the amount of information contained in these carriers is evaluated by a certain measure, it does not necessarily behave quantitatively with respect to the number of identically copied quantum carriers.

A question at this point is whether non-quantitative information (NQI) can be exhibited without employing particular measures of information.
Even if carrier $\ensrhn$ contains more information than $\enshfe$ does in a certain measure, it does not necessarily in an other measure \cite{JoszaSchlienz2000,Chefles2002}.
If a certain measure behaves non-quantitatively on a pair of carriers under copying, and if another measure evaluates their information content without copies differently, we say the pair exhibits {\it weakly} non-quantitative information (wNQI).
The choice of measure is essential for wNQI.

Pairs of quantum carriers, if carefully chosen, may exhibit NQI independently to the measure evaluating information content of carriers without their copies.
As we will show in the following, there are carriers $\ensrhn$ and $\enshfe$ such that the former alone offers better information about the system in any measure, but with copies, the latter performs better in a certain measure.
In contrast to wNQI, the measure only needs to be chosen on copied carriers, hence we say these carriers exhibit {\it strongly} non-quantitative information (sNQI) in this case.
A quantum carrier less informatic about the underlying physical system than another carrier on its own in any measure may still hide knowledge on the system and outperform the other one when multiple copies of them are compared.

The sNQI is not demonstrated in classical information theory, and it breaks the intuition that information content is a quantity inherent to its carriers.
Besides its fundamental interest, further analysis on the pair of carriers exhibiting the sNQI leads to observations on quantum information theory that wNQI does not.
Among the many observations we address three within this paper:quantum non-Markovianity exhibited by multiple uses of same channel sequences, incompleteness of what we call ``single-carrier'' measures, and a relationship between quantum information and hidden classical information potentially activated by copying.

This paper is constructed as follows:
We give definitions of carriers and classical information content in Sec.~\ref{sec:scmeasure}.
Based on these definitions we mathematically clarify what we mean by NQI.
The example of quantum carriers exhibiting sNQI is presented in Sec.~\ref{sec:snqi}, together with the proof of sNQI.
While sNQI is shown by using averaged fidelity as a measure of information content in Sec.~\ref{sec:snqi}, we compare averaged fidelity and mutual information in Sec.~\ref{sec:mutinfo}, to support the analysis of that averaged fidelity.
In Sec.~\ref{sec:discussion} we discuss the origin of sNQI, its relation to quantum non-Markovianity, and incompleteness of single-carrier measures.
In that discussion we reveal the existence of hidden classical information in quantum carriers.
Finally we conclude in Sec.~\ref{sec:conclusion} with some potential future applications.

\section{Single-carrier measures and non-quantitative information}\label{sec:scmeasure}
To explain the NQI precisely, we employ the following abstract treatments of quantum carriers, their information content, and strategies to obtain the information.

A quantum carrier refers to any physical system whose state is described by a density operator on a Hilbert space $\hil$.
The density operator of the carrier is assumed to be parameterized by a random variable $x \in X$ and denoted by $\rho_x \in \B(\hil)$; ($\B(\hil)$ denotes the space of linear operators on $\hil$).
Since a carrier investigated in this article is completely characterized by the ensemble $\E_\rho = \{ \rho_x, p_x \}_{x \in X}$ of quantum states with probability $p_x$ of the random variable, we sometimes use the symbol $\E_\rho$ to refer also to the corresponding carrier.

When copies of the carrier is not available, the observer gets a supply of single carriers in state $\rho_x$ with given probability $p_x$, on which they perform a measurement represented by positive operator-valued measure (POVM) elements $\{ E_y \in \B(\hil) \}_{y \in Y}$.
They obtain result $y$ with probability $p(y|x) = \trace[\rho_x E_y]$ and guess the value $x$ from $y$.
The guessing process is represented by the function $g:Y \rightarrow X,~y \mapsto g_y$.

When copies of the carrier are available, the observer gets a supply of two carriers in the same state $\rho_x \otimes \rho_x$ with probability $p_x$, on which they perform a joint measurement.
This is represented by POVM elements $\{ E_y \in \B(\hil \otimes \hil) \}_{y \in Y}$.
The observer obtains result for $y$ with the probability of $p(y|x) = \trace[\rho_x \otimes \rho_x E_y]$ and guess the value of $x$ from $y$.

A strategy by the observer is composed of the POVM measurement and the function of the guessing process.
The observer is assumed to be capable of performing any POVM measurement without experimental and statistical errors that cause shifts of the measurement probabilities from their true values.
The observer can optimize the strategy according to how density operators are parameterized by random variables and according to how the accuracy of guesses are estimated.

Measures of information content are real-valued functions of ensembles.
Since we consider information content obtainable by measurement strategies, its measures are real-valued functions of only measurement probabilities of single POVM measurements applied on carriers.
For the function $\ssm$ to be a measure of information content obtainable without copies it must satisfy the following:
Let $\E_\tau = \{ \tau_x \in \B(\hil_1), p_x \}_{x \in  X}$ and $\E_\rho = \{ \rho_x \in \B(\hil_2), p_x \}_{x \in  X}$ be ensembles with the same random variable.
If for any set of POVM elements $\{ E_y \in \B(\hil_1) \}_{y \in Y}$ there is a set of POVM elements $\{ E'_y \in \B(\hil_2) \}_{y \in Y}$ such that $\trace [ E_y \tau_x ] = \trace[ E'_y \rho_x ]$ holds for any $y \in Y$ and $x \in X$, then $\ssm (\E_\tau) \leq \ssm (\E_\rho)$.
In other words, if measurement results for ensemble $\E_\tau$ can be reproduced by measurement results for $\E_\rho$, information content of $\E_\tau$ must be estimated to be lower than or equal to that of $\E_\rho$.
Conversely, any function of probabilities obtained by single POVM measurements with the above described condition is regarded as a measure of information content obtainable without copies, and we call them single-carrier (SC) measures.

The set of SC measures thus defined contains distinguishability measures such as maximum probabilities of correct hypothesis testing \cite{Chefles1998hypothesis} and unambiguous state discrimination \cite{Helstrom1976}.
These measures include maximization or minimization with regard to measurement probabilities on single carriers in their definition.
If any measurement on $\enshfe$ can be simulated by those on $\ensrhn$, $\ensrhn$'s distinguishability should be evaluated higher.
This is because $\ensrhn$ has larger family of measurement probabilities over which optimization is taken.
There are also SC measures such as accessible information \cite{Holevo1973} and guessing probability \cite{YuenKennedyLax1975} which are not considered as distinguishability measures.

When an SC measure $\ssm$ is used to estimate information content of $\E_\rho = \{ \rho_x ,p_x \}_{x \in X}$ without copies, the corresponding measure of information content obtainable with the aid of single copy is $\ssm_\sr (\E_\rho) := \ssm (\{ \rho_x \otimes \rho_x , p_x \}_{x \in X})$.
Therefore we call $\ssm_\sr$ a double-carrier (DC) measure.
While both SC and DC measures are functions of probabilities obtained by a single POVM measurement, POVM measurements for DC measure may be jointly performed on the 2-copies of the same state from the ensemble.

NQI can be stated in a precise manner based on the presented setup.
When a pair of quantum carriers, $\E_\rho = \{ \rho_x , p_x \}_{x \in X}$ and $\E_\tau = \{ \tau_x , p_x \}_{x \in X}$,
satisfies the following two conditions:
\begin{eqnarray}
\label{eq:cond1}	\ssm (\E_\rho) &>& \ssm (\E_\tau),\\
\label{eq:cond2}	\ssm_\sr (\E_\rho) &<& \ssm_\sr (\E_\tau),
\end{eqnarray}
for an SC measure of $\ssm$, the pair is said to exhibit wNQI.
If the pair further satisfies
\begin{eqnarray}
\label{eq:cond3}	\ssm' (\E_\rho) \geq \ssm' (\E_\tau),
\end{eqnarray}
for any SC measure of $\ssm'$, the pair is said to exhibit sNQI.
In what follows we present a SC measure and an example pair of carriers exhibiting sNQI.

\section{exhibiting strongly non-quantitative information}\label{sec:snqi}
The random variable in this article is the vector $\vecn$ uniformly distributing over unit sphere $S_2$, which is called ``spin direction'' because of its relevance to particle physics.
The process of guessing $\vecn$ is also called ``orienteering'' \cite{Tangetal2020}.
For this random variable averaged fidelity used in \cite{Massar1995,GisinPopescu1999,Massar2000} estimates information content of a carrier.
Averaged fidelity $\fidelity (\E_\rho)$ as a SC measure for a carrier $\E_\rho =\{ \rho_\vecn, \dif \vecn \}_\sphere$ (here, $\dif \vecn$ represents the probability density for uniform distribution over unit sphere) is
\begin{eqnarray}
\label{eq:maxfidelity}	\fidelity (\E_\rho) := \max \int p(y|\vecn) \frac{1+\vecn \cdot \mathbf{g}_y}{2}  \dif \vecn \dif y,
\end{eqnarray}
where the maximization is over strategies constituted of POVM elements $\{ E_y \}_{y \in Y}$ and guessing process $g: y \mapsto \mathbf{g}_y \in S_2$.
The averaged fidelity increases as the guess direction $\mathbf{g}_y$ approaches the given direction $\vecn$ on average.
More precisely, the averaged fidelity estimates how much on average the observer can learn about the direction $\vecn$ from a given carrier with state $\rho_\vecn$, where the score of learning is $\cos^2 (\alpha/2) = (1+\vecn \cdot \mathbf{g}_y)/2$ with $\alpha$ being the angle between $\vecn$ and guess $\mathbf{g}_y$.
Maximizing $f$ is equivalent to minimizing the error of measurement strategy defined by the average of $\sin^2 (\alpha/2)$.

Dimensions of Hilbert spaces for our carriers $\ensrhn  = \{ \rhn , \dif \vecn \}_\sphere$ and $\enshfe = \{ \hfe , \dif \vecn \}_\sphere$ exhibiting sNQI are $2$ and $4$, respectively.
For later convenience we denote the Hilbert space for $\rhn$ by $\hil$ and that for $\hfe$ by $\hil \otimes \hil'$ where $\dim \hil = \dim \hil' =2$.
The density operators $\rhn$ and $\hfe$ are given by
\begin{eqnarray}
\label{eq:rhn}	\rhn &:=& \frac{\id_\hil + \sum_{i=1}^{3} n_i \sigma_i}{2}, \\
\label{eq:hfe}	\hfe &:=& \rhne \otimes \frac{\ketbra[0]}{2} + \rhmne \otimes \frac{\ketbra[1]}{2},
\end{eqnarray}
where $\id_\hil$ is the identity operator on $\hil$, $\sigma_i$ ($i=1,2,3$) are unitary Pauli operators, $\ket[0], \ket[1] \in \hil'$ are orthonormal vectors, and state $\rhne$ is defined by
\begin{eqnarray}
\label{eq:rhne}	\rhne = (1- \delta) \rhn + \delta \frac{\id_\hil}{2}
\end{eqnarray}
with a constant $\delta \in [0,1]$.

To check that carriers $\ensrhn$ and $\enshfe$  exhibit wNQI, we list the averaged fidelity for both carriers in TABLE~\ref{table}.
\begin{table}
\caption{\label{table}The averaged fidelity for carriers $\ensrhn$ and $\enshfe$ with and without their copies.
	 	Only lower bound is derived for $\fidelity_\sr(\enshfe)$ (``l.b.'' stands for lower bound).}
\begin{ruledtabular}
\begin{tabular}{ccc}
	 averaged fidelity & $\ensrhn$ & $\enshfe$ \\ \hline
	 without copies $\fidelity$ & $\frac{2}{3}$ & $\frac{2}{3} - \frac{\delta}{6}$\footnotemark[1] \\
	 with a single copy $\fidelity_\sr$ & $\frac{3}{4}$\footnotemark[2] & l.b.: $ \frac{2\sqrt{3} + 15}{24} - \frac{2 \sqrt{3} + 3}{24} \delta$\footnotemark[1] \\
\end{tabular}
\footnotetext[1]{See Appendix~\ref{sec:fidelity}.}
\footnotetext[2]{Reference \cite{Massar1995}}
\end{ruledtabular}
\end{table}
When $\delta =0$, ensembles $\ensrhn$ and $\enshfe$ have the same averaged fidelity, and $\fidelity(\enshfe)$ decreases as $\delta$ increases.
Especially condition (\ref{eq:cond1}) is satisfied for non-zero $\delta$.
While $\fidelity_\sr(\enshfe)$ has not been obtained, we have constructed a strategy $(\{ E_y \}_{y\in Y}, g)$ giving its lower bound $(2\sqrt{3} + 15)/24 - (2 \sqrt{3} + 3)\delta/24$, which is greater than $\fidelity_\sr(\ensrhn) = 3/4$ when $\delta \leq 7-4\sqrt{3} \approx 0.0718$.
That is, carriers $\ensrhn$ and $\enshfe$ exhibits wNQI when $0 < \delta < 7-4\sqrt{3}$.

\begin{figure}[htbp]
	\includegraphics[width=8.6cm]{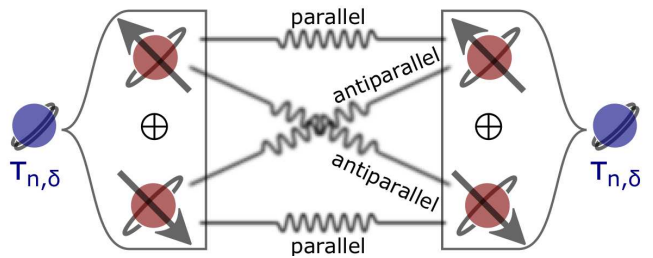}
	\caption{\label{fig:strategy} Measurement strategy to obtain the lower bound of the averaged fidelity for two copies of carrier $\enshfe$. Parallel and antiparallel spin pairs are measured by different entangled POVM elements.}
\end{figure}

We make a rough sketch of the measurement strategy for obtaining the fidelity $(2 \sqrt{3} + 15)/24 - (2 \sqrt{3}+3)\delta/24 $ from two copies of carrier $\enshfe$ (see Appendix~\ref{sec:lowerboundfidelity} for details).
The state $\hfe \otimes \hfe$ is a direct sum of noisy parallel and antiparallel spins.
We employ strategies for parallel spins presented in \cite{Massar1995,Massar2000} and for antiparallel spins presented in \cite{GisinPopescu1999,Massar2000} (as depicted in Fig.~\ref{fig:strategy}).
This strategy for $\hfe \otimes \hfe$ outperforms the optimal strategy for parallel spins since antiparallel spins offer better averaged fidelity than parallel ones as theoretically shown in \cite{GisinPopescu1999,Massar2000}.

To show sNQI, it remains to prove condition (\ref{eq:cond3}).
In Appendix~\ref{sec:stat} we construct a unital positive map $\ptp_\delta:\B(\hil \otimes \hil') \rightarrow \B(\hil)$ such that
\begin{eqnarray}
\label{eq:unitalmap}	\trace[ E \hfe ] = \trace[ \ptp_\delta (E) \rhn] ~ (\forall \vecn \in S_2),
\end{eqnarray}
for any operator $E \in \B(\hil \otimes \hil')$.
Importantly, it is proven with the aid of theorems from \cite{CheflesJoszaWinter2004} that the map $\ptp_\delta$ satisfying Eq.~(\ref{eq:unitalmap}) cannot be completely positive.
Existence of the map $\ptp_\delta$ satisfying Eq.~(\ref{eq:unitalmap}) is sufficient for condition (\ref{eq:cond3}).
In fact any POVM measurement with elements $\{ E_i \}_{i \in I}$ on ensemble $\enshfe$ is simulated by the POVM measurement with elements $\{ \ptp_\delta (E_i) \}_{i \in I}$ on $\ensrhn$.

Remarkably, condition (\ref{eq:cond3}) is satisfied with equality for any SC measure at $\delta = 0$.
This can be observed by the inverse of relation (\ref{eq:unitalmap}), namely, there is a unital positive map $\mathcal{J}:\B(\hil) \rightarrow \B(\hil \otimes \hil')$ such that
\begin{eqnarray}
\label{eq:unitalmap2}	\trace[ \mathcal{J}(E) \tau_{\vecn,0} ] = \trace[ E \rhn] ~ (\forall \vecn \in S_2),
\end{eqnarray}
for any operator $E \in \B(\hil)$ (see Appendix~\ref{sec:stat}).
Any SC measure is evaluated to be same for ensembles $\ensrhn$ and $\enshfe$ at $\delta=0$, since any POVM measurement on carrier $\ensrhn$ can be simulated by that on $\enshfe$ and vice versa.

In summary, the pair of carriers $\ensrhn$ and $\enshfe$ whose states defined by Eqs.~(\ref{eq:rhn}) and (\ref{eq:hfe}), satisfies conditions (\ref{eq:cond1}), (\ref{eq:cond2}) and (\ref{eq:cond3}) when $0 < \delta < 7-4\sqrt{3}$.
These carriers exhibit sNQI: classical information content of these carriers reverses when copies are available.
Without copies, the spin direction cannot be guessed more accurately by measurements on $\enshfe$ than on $\ensrhn$ in any figure of merit.
With copies, that is, when pairs of these carriers are compared, the averaged fidelity of $\enshfe$ is higher than that of $\ensrhn$.
Carriers with poor information content in any figure of merit may be very informatic with the aid of their copy in some figures of merit.

\section{comparison with mutual information}\label{sec:mutinfo}
\begin{figure}[htbp]
	\includegraphics[width=8.6cm]{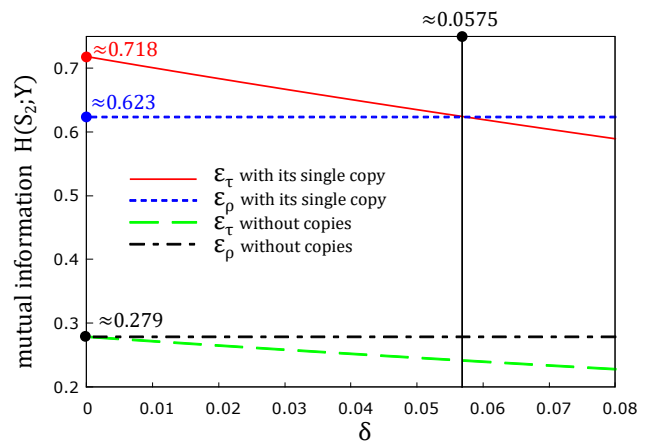}
	\caption{\label{fig:mutinfo} Mutual information (\ref{eq:mutinfo}) between spin direction and observers' register obtained by measuring carriers $\ensrhn$ and $\enshfe$ with and without their copies. The POVM elements $\{ E_y \}_{y \in Y}$ of the observers' measurement are those we used to obtain the values of fidelity listed in TABLE~\ref{table}. Mutual information for $\enshfe$ with its single copyt is higher than that for $\ensrhn$ when $0 \leq \delta \lessapprox 0.0575 $. See Appendix~\ref{sec:2} for derivations and analytic forms of these mutual information.}
\end{figure}

averaged fidelities calculated above for showing sNQI does not contradict values of mutual information.
In FIG.~\ref{fig:mutinfo}, we plot mutual information
\begin{eqnarray}
\label{eq:mutinfo}	H (S_2;Y) := \int p(y|\vecn) \log \frac{p(y|\vecn)}{p(y)}  \dif \vecn \dif y,
\end{eqnarray}
between spin direction $S_2$ of the underlying physical system and observers' register $Y$ created by the measurements giving fidelities listed in TABLE.~\ref{table}.
With a single copy, mutual information of $\enshfe$ is larger than that of $\ensrhn$ for small enough $\delta$.
Averaged fidelity and mutual information have an overlap region of $\delta$ in which the order of their value is reversed under copying.

Currently we are not sure if the carriers $\ensrhn$ and $\enshfe$ exhibit sNQI with accessible information, namely, the maximally attainable mutual information.
Under an assumption that the optimal strategy constitutes of covariant measurements \cite{Holevo1982}, the values of mutual information plotted in FIG.~\ref{fig:mutinfo} for $\E_\rho$ with and without its copies, and for $\E_\tau$ without its copies are maximum (see Appendix~\ref{sec:2} for the proof).
Accessible information demonstrates sNQI if the assumption is true.

\section{discussion and implications}\label{sec:discussion}
\subsection{Non-positivity of doubled statistical morphisms}\label{sec:statistical}
Perhaps sNQI goes against one's intuition if one knows classical information theory because it is never demonstrated by any pair of probabilistic carriers, of which states are described by random variables.
In terms of the difference between quantum and probabilistic carriers, sNQI originates from the gap between completely positive maps and statistical morphisms \cite{Buscemi2012,Buscemi2016}.

To understand the origin of sNQI, it is helpful to review why probabilistic carriers never demonstrate sNQI.
Our abstract treatments of physical systems can be restricted to probabilistic carriers by assuming simultaneous diagonalizability of all operators in a fixed basis: states and POVM operators on a genreral probabilistic carrier can be decomposed to the form $\sum_i r_i \ketbra[i]$ with non-negative numbers $r_i$ and a common basis $\{ \ket[i] \}_i$.
Condition (\ref{eq:cond3}) of sNQI implies the existence of a classical (dual) stochastic map $L$ such that $\trace[E \tau_x] = \trace[L(E) \rho_x]$ for all $x \in X$ and all POVM operator $E$ for probabilistic carriers (this may seem obvious; see Appendix~\ref{sec:classical} for the proof).
For any set of POVM elements $\{ E_y \}_{y \in Y}$ on carrier $\{ \tau_x \otimes \tau_x, p_x \}$ we have
\begin{eqnarray}
\label{eq:tensor_statistical_morphisms}	\trace [ E_y \tau_x \otimes \tau_x] = \trace [ L \otimes L  (E_y) \rho_x \otimes \rho_x],
\end{eqnarray}
for all $x \in X$.
Since $\{ L \otimes L (E_y) \}_{y \in Y}$ is a valid set of POVM elements on carrier $\{ \rho_x \otimes \rho_x, p_x \}$, Eq.~(\ref{eq:tensor_statistical_morphisms}) reveals that any measurements on $\{ \tau_x \otimes \tau_x, p_x \}$ can be simulated by appropriate measurements on $\{ \rho_x \otimes \rho_x, p_x \}$.
Thus we have shown that condition (\ref{eq:cond3}) implies
\begin{eqnarray*}
	\ssm_\sr (\ensrhn) \geq \ssm_\sr (\ensrhn) ~(\forall \ssm_\sr:\text{DC measure}),
\end{eqnarray*}
for probabilistic carriers, from which we observe the ``quantitatibity of classical information'' in classical information theory.

Equation (\ref{eq:unitalmap}) for our quantum case and $\trace[E \tau_x] = \trace[L(E) \rho_x]$ for the classical case state that $\ptp_\delta$ and $L$ are statistical morphisms \cite{Buscemi2012,Buscemi2016} from the system of $\enshfe$ to $\ensrhn$.
These two statistical morphisms both offer prescriptions of how to simulate measurements on $\enshfe$ by those on $\ensrhn$.
While classical statistical morphisms such as $L$ are automatically completely positive, quantum statistical morphisms such as $\ptp_\delta$ are not necessarily.

The unital positive map $\ptp_\delta$ satisfying Eq.~(\ref{eq:unitalmap}) is not completely positive and the parallel application of two maps $\ptp_\delta \otimes \ptp_\delta$ is no more positive.
Hence, even though we have
\begin{eqnarray*}
	\trace[ E \hfe \otimes \hfe ] = \trace[ \ptp_\delta \otimes \ptp_\delta (E) \rhn \otimes \rhn] ~ (\forall \vecn \in S_2),
\end{eqnarray*}
similarly to Eq.~(\ref{eq:tensor_statistical_morphisms}), POVM measurement with element $E$ on $\hfe \otimes \hfe$ is not necessarily simulated by element $\ptp_\delta \otimes \ptp_\delta (E)$ on $\rhn \otimes \rhn$ because $\ptp_\delta \otimes \ptp_\delta (E)$ may not be valid POVM operator for entangled $E$.
Our measurement strategy on $\{ \hfe \otimes \hfe, \dif \vecn \}_\sphere$ depicted in Fig.~\ref{fig:strategy} makes use of such entangled POVM measurements to exhibit sNQI.

Gisin and Popescu showed in \cite{GisinPopescu1999} that antiparallel spins $\rhn \otimes \rhmn$ are more informatic than parallel spins $\rhn \otimes \rhn$ about the spin direction $\vecn$.
They emphasize that parallel and antiparallel spins produce different results since the ``spin-flip'' operation $\mathrm{SFlip}:\rhn \mapsto \rhmn$ is not completely positive and operation $\mathrm{Id} \otimes \mathrm{SFlip}: \rhn \otimes \rhn \leftrightarrow \rhn \otimes \rhmn$ ($\mathrm{Id}$ represents the identity operation) is not realizable even passively because it is not positive.
Initially, we could intuitively suppose that the antiparallel spins $\rhn$ and $\rhmn$ compensate the knowledge of spin direction of each other to have superiority over parallel spins.

Gisin and Popescu's observation about the non-completely positive maps also applies to sNQI, except that sNQI requires non-positivity of doubled maps such as $\ptp_\delta \otimes \ptp_\delta$, rather than combination of different positive maps such as $\mathrm{SFlip}$ and $\mathrm{Id}$.
Thanks to the use of non-positive doubled maps, sNQI disproves the intuitive idea that two carriers compensate their knowledge of each other.
Rather, sNQI reveals that some carriers are more ``self-cooperative'' than others.

\subsection{Quantum non-Markovianity exhibited by doubled channel sequences}
The difference between Markov processes in classical and quantum information theory (see \cite{RivasHuelgaPlenio2014:markov,BreuerLainePiiloVacchini2016} for review) is highlighted by sNQI.
As is noted in Sec.~\ref{sec:statistical}, the phenomenon of sNQI emerges because the statistical morphism $\ptp_\delta$ is not completely positive.
Quantum non-Markovianity, which here means increase of information according to quantum stochastic processes \cite{Breueretal2009,Chruscinskietal2011,BuscemiDatta2016,BaeChruscinski2016}, also originates from non-completely-positive reduced-state transformations.
Not only do these two phenomena have a common origin, but sNQI can be regarded as an exotic kind of quantum non-Markovianity phenomenon.

\begin{figure}[htbp]
	\includegraphics[width=8.6cm]{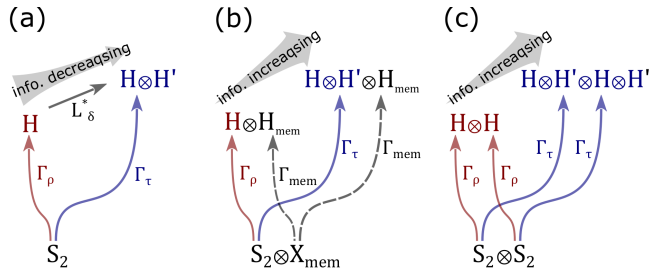}
	\caption{\label{fig:markov} Evolution of information content along the sequences of channels. (a) The existence of statistical transformation $\ptp_\delta$ ensures information only decreases along the sequence $(\Gamma_\rho,~\Gamma_\tau)$ without a memory. (b) Non-complete positivity of $\ptp_\delta$ reveals the existence of a memory channel $\Gamma_{mem}$ with which information content may increase along the sequence $(\Gamma_\rho \otimes \Gamma_{mem},~\Gamma_\tau \otimes \Gamma_{mem})$ for proper inputs. (c) Information content is shown to increase along the doubled sequence $(\Gamma_\rho \otimes \Gamma_\rho,~\Gamma_\tau \otimes \Gamma_\tau)$ by sNQI.}
\end{figure}

Let us consider a sequence of classical-input quantum-output channels $(\Gamma_\rho: S_2 \rightarrow \B(\hil),~\Gamma_\tau: S_2 \rightarrow \B(\hil \otimes \hil'))$ defined by $\Gamma_\rho (\vecn) = \rhn$ and $\Gamma_\tau (\vecn) = \hfe$.
The existence of positive map $\ptp_\delta$ implies Markovianity of sequence $(\Gamma_\rho,~\Gamma_\tau)$ in any of its classical snap-shots:
for any POVM measurement $\{ E_j \}_{j \in J}$ on $\enshfe$ there exists a POVM measurement $\{ F_j \}_{j \in J}$ on $\ensrhn$ such that $\trace[ E_j \hfe] = \trace[ F_j \rhn]$ holds for any $\sphere$.
Information content decreases according to the channel sequence when the sequence itself does not accompany another physical system as a memory, as is depicted in Fig.~\ref{fig:markov} (a).

Nevertheless, sequence $(\Gamma_\rho,~\Gamma_\tau)$ exhibits quantum non-Markovianity by taking the memory into account.
According to corollary 5 of \cite{Buscemi2016}, if $\ptp_\delta$ is not completely positive, there is a triad of a random variable $X_{mem}$, Hilbert space $\hil_{mem}$ and a third channel $\Gamma_{mem} :X_{mem} \rightarrow \B(\hil_{mem})$, such that information content measured by guessing probability increases along the channel sequence $(\Gamma_\rho \otimes \Gamma_{mem},~\Gamma_\tau \otimes \Gamma_{mem})$ for an appropriate input (see Fig.~\ref{fig:markov} (b) for a schematic representation).
This is a standard analysis employed for characterizing information flow for quantum non-Markovianity phenomena, where the third channel $\Gamma_{mem}$ plays the role of a memory \cite{Buscemi2016,BuscemiDatta2016,BaeChruscinski2016}.

Here, as is depicted in Fig.\ref{fig:markov} (c), sNQI tells us that the increase of information content is simply demonstrated by the doubled sequence $(\Gamma_\rho \otimes \Gamma_\rho,~\Gamma_\tau \otimes \Gamma_\tau)$.
One can make a more accurate guess on unit vector $\vecn$ from state $\Gamma_\tau \otimes \Gamma_\tau (\vecn, \vecn) = \hfe \otimes \hfe$ than from state $\Gamma_\rho \otimes \Gamma_\rho (\vecn , \vecn) = \rhn \otimes \rhn$ by inputting $(\vecn, \vecn )$ to the doubled sequence.
If one does not understand sNQI, one would doubt that information content ever increases along the doubled sequence $(\Gamma_\rho \otimes \Gamma_\rho,~\Gamma_\tau \otimes \Gamma_\tau)$, since it is a combination of two channel sequences which both have the property of information-decreasing.
The use of a doubled channel instead of the third memory channel makes a sharp contrast to standard quantum non-Markovianity analysis.

Even if a channel sequence is divided by statistical morphisms that are not completely positive, its doubled sequence does not necessarily exhibit increase of information content.
An obvious example of statistical morphism is transposition, which does not cause sNQI since the parallel action of transpositions results again in transposition.
Thus, it is more challenging to exhibit quantum non-Markovianity by sNQI than by the aid of memory channels.

\subsection{Hidden information lead from incompleteness of SC measures}\label{sec:incompleteness}
Comparison of carriers $\ensrhn$ and $\enshfe$ in their information content leads to a consequence on quantum statistics which we call {\it incompleteness} of SC measures.
The set of all SC measures on ensembles of classical probability distributions is sufficient for identifying the ensembles' information content completely, because any measure is SC for these ensembles.
According to sNQI, however, there is hidden information in ensembles of quantum states which cannot be witnessed by any SC measure.
Moreover, even if the values of all SC measures are available at the same time, one cannot recognize the information hidden in $\enshfe$ potentially activated by copying.
In fact, at $\delta =0 $, any SC measure is evaluated to the same for $\ensrhn$ and $\enshfe$, while at least one DC measure is evaluated higher for $\enshfe$.
In this sense the set of all SC measures is incomplete among all measures, since they are not sufficient for recognizing the hidden information potentially activated by copying.

If SC measures do not witness the hidden information potentially activated by copying, which measure effectively detects it without the use of measurements on copied systems?
The incompleteness of SC measures tells us that such a measure does not estimate {\it classical} information extracted by measurements.
Therefore, it is worth comparing the DC measures and measures of {\it quantum} information to see if classical information is hidden in a form of quantum information.
While the notion of quantum information itself is ambiguous \cite{Josza2004}, there are several functions considered to measure quantum information of ensembles.
We consider here two representative and calculable functions of quantum information: Holevo $\chi$ quantity \cite{Holevo1973} and the optimal blind compression rate \cite{JoszaSchumacher1994,Schumacher1995,KoashiImoto2001:compressibility}.

Holevo quantity $\chi$ \cite{Holevo1973} is an example that is neither SC nor a witness of hidden information.
We have
\begin{eqnarray*}
	\chi(\ensrhn) = 1,
\end{eqnarray*}
and
\begin{eqnarray*}
	\chi(\enshfe) = 1 + \left( 1-\frac{\delta}{2} \right) \log \left( 1-\frac{\delta}{2} \right) + \frac{\delta}{2} \log \frac{\delta}{2}.
\end{eqnarray*}
The latter is a monotonically decreasing function of $\delta$ taking the maximum value $1$ at $\delta=0$.
Thus the Holevo $\chi$ quantity is not evaluated higher for $\enshfe$.

Optimal compression rate $R$ of blind compression task, in which the message sender has to compress a sequence of unidentified quantum states supplied from a source, is evaluated higher for ensemble $\enshfe$ than for $\ensrhn$.
It is given by von Neumann entropy for ensembles constituted only of pure states such as $\ensrhn$ \cite{JoszaSchumacher1994,Schumacher1995}, and can be calculated according to the prescription from \cite{KoashiImoto2001:compressibility,KoashiImoto2002:operations} for ensembles of general mixed states.
We have
\begin{eqnarray*}
	R(\ensrhn) = 1,
\end{eqnarray*}
and
\begin{eqnarray*}
	R(\enshfe) = \left\{ \begin{array}{lc}
	2 & (0 \leq \delta <1) \\
	0 & (\delta =1).
	\end{array}\right.
\end{eqnarray*}
Thus the optimal blind compression rate witnesses the hidden information contained in $\enshfe$.

This result on blind compression rate, together with the incompleteness of SC measures, extends a discrepancy between von Neumann entropy and pairwise fidelity found in \cite{JoszaSchlienz2000} and called ``Jozsa-Schlienz paradox'' in \cite{Yang2005}.
We have $\fidelity_p (\rhn, \rho_{\vect[m]}) = \fidelity_p (\tau_{\vect[n],0}, \tau_{\vect[m],0})$ for all the pairs of unit vectors $\vecn,\vect[m] \in S_2$, for pairwise fidelity $\fidelity_p(\rho_1,\rho_2) := \trace (\rho_1^{1/2} \rho_2 \rho_1^{1/2})^{1/2}$.
However, the blind compression rate of ensembles $\ensrhn$ and $\enshfe$ at $\delta=0$ differ.
Thus, it is possible to change the blind compression rate while keeping the values of all pairwise fidelity and all SC measures of the ensemble.
The same discrepancy is previously known between von Neumann entropy and pairwise fidelity for pure state ensembles \cite{JoszaSchlienz2000}.
Here we extend the discrepancy to mixed state ensembles where von Neumann entropy is generalized to blind compression rate, and under this generalized setting, answer to a question remain opened in \cite{JoszaSchlienz2000}: SC measures such as accessible information and minimum error probability does not help calculating blind compression rate for mixed state ensembles.

Even if one can interpret the hidden information as a form of quantum information conveyed by the single carrier, the impossibility to extract the hidden information without the aid of carriers' copy is not reversed.
Thus, the presented sNQI reveals that the concept of ``classical information'' is independent of its ``carrier'' in quantum theory.
When we say ``classical information is conveyed by its carrier,'' it is assumed that the carrier itself has an inherent ability to convey the information.
It is already known that this inherent ability does not behave perfectly quantitatively when different carriers are combined \cite{GisinPopescu1999}, and that there are physical realization of carriers not based on particles transmitted from senders to receivers \cite{Salihetal2013,Arvidsson-ShukurBarnes2016,Arvidsson-Shukuretal2017,Caoetal2017,Salihetal2018,AharonovVaidman2019,Vaidman2019,Calafelletal2019}.
Still, we intuitively consider good carriers remain good when same copies of them can be used at the same time.
This intuition lasting in a small way finally collapses by the discovery of sNQI.
If carriers ``contain'' classical information, where is the carrier containing hidden information?
Classical information requires a carrier when it is conveyed.
However, the ability to convey classical information is not inherent in each carrier, but in the final structure of carriers at the message receiver.

\section{Conclusion and perspectives}\label{sec:conclusion}
It is theoretically demonstrated that the classical information content of quantum carriers does not necessarily grow quantitatively along the number of simultaneously available carrier copies.
To show this, we first formulate SC measures as estimators of classical information contained by the single carrier.
DC measures, which estimate classical information contained by two copies of the same carrier, simply correspond to SC measures on two copies of carriers regarded as a single carrier.
Then we give an example of two quantum carriers such that all SC measures record at the first carrier higher than or equal to the second carrier, and that at least one DC measure is evaluated higher for the second carrier.
These carriers are said to exhibit ``strongly'' non-quantitative information because the superiority of the single first carrier does not depend on the SC measure.

The state of our first carrier $\ensrhn$ is in usual spin-$1/2$ particle, parameterized by the spin direction.
The state of our second carrier $\enshfe$ is a direct sum of two noisy spins, where the spin direction of the half side is flipped.
Any SC measures reveal that the first carrier alone is more useful than the second alone for orienteering the spin direction.
However, averaged fidelity is higher for two copies of the second carriers than for two copies of the first ones and values of mutual information are in accord to the averaged fidelity.

Our method to show sNQI relies on Gisin and Popescu's result \cite{GisinPopescu1999} that antiparallel spins are more useful than parallel spins for orienteering.
After their finding, additional results followed which detailed the advantage of antiparallel spins; for example, in the results of analyses in figures of merit other than the averaged fidelity \cite{Changetal2014}, of orienteering spin direction distributed on circles \cite{BraunsteinGhoshSeverini2007}, and in the advantage of conjugate copies for quantum cloning \cite{CerfIblisdir2001,Kato2009}.
It would be interesting to adjust these analyses to our carriers to uncover deeper understanding of sNQI.

The sNQI emerges from the gap between statistical morphisms and completely positive maps---which only exists in quantum theory.
More precisely, sNQI is made possible by the fact that a parallel application of two identical positive maps does not necessarily results in a valid (passive) transformation.
After clarifying this origin of sNQI, we interpret sNQI as a kind of quantum non-Markovianity phenomenon.

To construct other examples of carriers exhibiting sNQI, the natural first step is to find statistical morphisms like $\ptp_\delta$ which become invalid when doubled $\ptp_\delta \otimes \ptp_\delta$.
Positive maps which are not copositive \cite{Choi1975,Choi1980,Choetal1992,Ha1998,Terhal2001} are candidates of statistical morphisms to be tested.
Particularly if a positive map $L$ remains positive until $N$ copies $L \otimes L \otimes ... \otimes L$ and does not from $N+1$ copies, it might be possible to construct two carriers so that the amount of classical information content reverses at $N+1$ copies.

The presented sNQI implies that the capability of quantum carriers to convey classical information cannot be completely specified from SC measures.
This incompleteness of SC measures reveals the existence of hidden information which is opened when carriers are available with their copies.
The optimal blind compression rate is shown to witness the hidden information for our example, from which we observe that the unavailable classical information is hidden in a form of quantum information.
Based on this result, we extend Jozsa-Shlienz's paradox \cite{JoszaSchlienz2000} to the mixed state regime, where von Neumann entropy is generalized to the blind compression rate and pairwise fidelity is enlarged for all SC measures.

Other measures of quantum information could also be tested as witnesses of hidden information.
For example, our analysis on Holevo quantity did not conclude that the optimal visible compression rate \cite{Horodecki1998:limits,Horodecki2000:purification,BarnumCavesFuchsJoszaSchumacher2001,Hayashi2006} witnesses the hidden information.
It is necessary to further investigate the relationship between the hidden information and quantum information to ascertain whether the sNQI phenomenon could be useful for information processing such as quantum cryptography.

Another direction that should be explored is the classical information of quantum state transformations, rather than states themselves.
The field of quantum metrology pertains to tasks which extract classical information from collective actions of the same quantum state transformations \cite{Giovanettietal2006,Giovanettietal2011}.
The analysis on hidden information of state transformations will therefore have direct consequences to the field. The results of quantum states do not necessarily apply to state transformations, as they behave differently under copying \cite{Chiribella2015} and complex conjugation \cite{Miyazaki2019}.

Finally, after over 20 years since Gisin and Popescu's theoretical prediction \cite{GisinPopescu1999}, an experiment confirmed that antiparallel spins outperform parallel spins for orienteering \cite{Tangetal2020}.
Therefore the author believes that experimental realization of our carriers would require technologies readily available at this time and hope that the sNQI will be experimentally demonstrated in the near future.

\begin{acknowledgments}
We thank E. Wakakuwa for helpful comments.
\end{acknowledgments}

\appendix

\section{Statistical morphisms}\label{sec:stat}
In this section, we construct unital positive maps $\ptp_\delta:\B(\hil \otimes \hil') \rightarrow \B(\hil)$ and $\mathcal{J}:\B(\hil) \rightarrow \B(\hil \otimes \hil')$ satisfying Eqs.~(\ref{eq:unitalmap}) and (\ref{eq:unitalmap2}), respectively.
These maps are examples of statistical morphisms studied in \cite{Buscemi2016}.
For later convenience we denote the ensemble $\{ \rhne , \dif \vecn \}_{\vecn \in S_2}$ by $\ensrhn'$.

We decompose $\ptp_\delta$ into a sequence $\dep_\delta \circ \ptp_0  $ of a statistical morphism $\ptp_0$ from $\enshfe$ to $\ensrhn'$ and the conjugate of depolarizing channel $\dep_\delta$.
First define $\ptp_0:\B(\hil \otimes \hil') \rightarrow \B(\hil)$ by
\begin{eqnarray}
\nonumber	\ptp_0 (E) &=& \frac{1}{2} \trace_{\hil'} [ E (\id_\hil \otimes \ketbra[0])] \\
\label{eq:def_ptp0}	&+& \frac{1}{2} \pauliy \left\{ \trace_{\hil'} \left[ E (\id_\hil \otimes \ketbra[1]) \right] \right\}^\text{T} \pauliy,
\end{eqnarray}
where the transposition $\text{T}$ is taken in the basis $\ket[0],\ket[1]$.
Then we have
\begin{eqnarray}
\nonumber	\trace[ \ptp_0 (E) && \rhne] \\
\nonumber	= \frac{1}{2} \trace && \left[\rhne \trace_{\hil'} [ E (\id_\hil \otimes \ketbra[0])] \right] \\
\nonumber	&&+ \frac{1}{2} \trace \left[ (\pauliy \rhne \pauliy)^\ast \trace_{\hil'} \left[ E (\id_\hil \otimes \ketbra[1]) \right] \right] \\
\nonumber	= \frac{1}{2} \trace && \left[\rhne \trace_{\hil'} [ E (\id_\hil \otimes \ketbra[0])] \right] \\
\nonumber	&&+ \frac{1}{2} \trace \left[ \rhmne \trace_{\hil'} \left[ E (\id_\hil \otimes \ketbra[1]) \right] \right] \\
\nonumber	= \frac{1}{2} \trace && \left[ E (\rhne \otimes \ketbra[0]) \right] + \frac{1}{2} \trace \left[ E (\rhmne \otimes \ketbra[1]) \right] \\
\label{eq:ptp0}	= \trace[ E && \hfe ],
\end{eqnarray}
where the second equality follows from
\begin{eqnarray*}
	(\pauliy \rhne \pauliy)^\ast &=& (1- \delta) \pauliy \rhn^\ast \pauliy + \delta \pauliy \left( \frac{\id_\hil}{2} \right)^\ast \pauliy\\
	 &=& (1-\delta) \rhmn + \delta \frac{\id_\hil}{2}.
\end{eqnarray*}
The linear map $\ptp_0$ is positive since it is a convex sum of two positive maps $E \mapsto \trace_{\hil'} [ E (\id_\hil \otimes \ketbra[0])]$ and $E \mapsto \pauliy \left\{ \trace_{\hil'} \left[ E (\id_\hil \otimes \ketbra[1]) \right] \right\}^\text{T} \pauliy$, where the latter half is not completely positive.

Since $\rhne$ is obtained from $\rhn$ through depolarizing channel $\rho \mapsto (1-\delta) \rho + \delta \id_\hil /2$ on $\hil$, the conjugation of depolarizing channel (in Hilbert-Schmidt inner product)
\begin{eqnarray*}
	&& \dep_\delta : \B (\hil) \rightarrow \B (\hil),\\
	&& \dep_\delta (E) := (1-\delta) E + \delta \id_\hil ~~(\forall E \in \B(\hil)),
\end{eqnarray*}
satisfies
\begin{eqnarray}
\label{eq:dep}	\trace [ \dep_\delta (E) \rhn ] = \trace [ E \rhne],
\end{eqnarray}
for any operator $E \in \B(\hil)$.
Combining equations (\ref{eq:ptp0}) and (\ref{eq:dep}) we have
\begin{eqnarray*}
	\trace [ \dep_\delta \circ \ptp_0 (E) \rhn ] = \trace[ \ptp_0 (E) \rhne ] = \trace[ E \hfe],
\end{eqnarray*}
for any operator $E \in \B(\hil)$ as required.

Now we turn to the inverse relation.
Define a unital positive map $\mathcal{J} : \B(\hil) \rightarrow \B(\hil \otimes \hil')$ by
\begin{eqnarray}
\label{eq:inverse_statistic}	\mathcal{J} (E) := E \otimes \ketbra[0] + \pauliy E^\text{T} \pauliy \otimes \ketbra[1].
\end{eqnarray}
Then we have
\begin{eqnarray}
\nonumber	\trace[ && \mathcal{J}(E) \hfe] \\
\nonumber &&= \frac{1}{2} \trace[ E \rhne \otimes \ketbra[0] ] + \frac{1}{2} \trace[ \pauliy E^\text{T} \pauliy \rhmne \otimes \ketbra[1] ] \\
\nonumber	&&= \frac{1}{2} \trace[ E \rhne \otimes \ketbra[0] ] + \frac{1}{2} \trace[ E \rhne \otimes \ketbra[1] ] \\
\label{eq:j}	&&= \trace[ E \rhne ]
\end{eqnarray}
for any Hermite operator $E \in \B(\hil)$ and $\vecn \in S_2$.
Equation (\ref{eq:j}) includes Eq.~(\ref{eq:unitalmap2}) as a special case where $\delta=0$.

While Eq.~(\ref{eq:ptp0}) implies that any measurement on carrier $\enshfe$ is simulated by a corresponding measurement on $\ensrhn'$, Eq.~(\ref{eq:j}) implies that the other way around is also true.
Thus information content of carriers $\enshfe$ and $\ensrhn'$ are estimated to be equal by any SC measures.
When calculating averaged fidelity and mutual information of $\enshfe$ in Secs.~\ref{sec:fidelity} and \ref{sec:2}, we use the fact that information content of $\enshfe$ and $\ensrhn'$ are equivalent in any SC measures.

\section{Averaged fidelity}\label{sec:fidelity}
\subsection{averaged fidelity of $\enshfe$ without copies}\label{sec:fidelity_singletau}
Instead of calculating the averaged fidelity $\fidelity(\enshfe)$ directly from $\enshfe$, we use equality
\begin{eqnarray*}
	\fidelity (\enshfe) = \fidelity ( \ensrhn' ),
\end{eqnarray*}
just noted at the end of Sec.~\ref{sec:stat}, and search the optimal measurement strategy for maximizing fidelity function $f$ of ensemble $\ensrhn'$.
The explicit form of fidelity function $f$ is
\begin{eqnarray*}
	f (\ensrhn', \{ E_y \}_{y \in Y}, g) = \int \dif \vecn \dif y \trace[ \rhne E_y ] \frac{1+\vecn \cdot \mathbf{g}_y}{2},
\end{eqnarray*}
where $\{ E_y \}_{y \in Y}$ is a elements and $g:Y \rightarrow S_2,~y \mapsto \mathbf{g}_y$ is a function.
We first show that the optimal measurement can be assumed covariant.
Then the form of POVM operators is reduced so that they depend only on the direction $\vecn$ and a single parameter.
The optimization is completed by maximizing the fidelity along the parameter.

Let $\{ E_y \}_{y \in Y}, g$ be an optimal strategy.
Without loss of generality, we can assume that the POVM elements are labeled by $\vect[m]$ in image of function $g$ because
\begin{eqnarray*}
	\int_{\mathbf{g}_y = \vect[m]} \dif y \trace[ \rhne E_y ] && \frac{1+\vecn \cdot \mathbf{g}_y}{2} \\
	&&= \trace \left[  \rhne \int_{\mathbf{g}_y = \vect[m]} \dif y E_y \right] \frac{1+\vecn \cdot \vect[m]}{2},
\end{eqnarray*}
implies that integrating POVM elements up to have $E_{\vect[m]} = \int_{\mathbf{g}_y = \vect[m]} \dif y E_y$ does not change the fidelity.
Since POVM elements are labeled by the spin direction which is to be the guess, and since the spin direction is uniformly distributing over the sphere, we can assume that the measurement is covariant.
That is, the label $\vect[m]$ spreads all over the sphere $S_2$, and we have
\begin{eqnarray*}
	E_{R(\vect[m])} = U_R E_{\vect[m]} U_R^\dagger.
\end{eqnarray*}
if $R$ is a rotation on $S_2$, and $U_R$ is its representation on $\hil$.

Let us consider the POVM operator $E_\uparrow$ for unit vector $\uparrow := (0,0,1)$.
Trace of $E_\uparrow$ is
\begin{eqnarray*}
	\trace[E_\uparrow] &=& \int \dif \vecn \trace [ U_{R_\vecn} E_\uparrow U_{R_\vecn}^\dagger ] \\
	&& = \trace [ \int \dif \vecn E_{R_\vecn(\uparrow)} ] = \trace[ \id_\hil] = 2,
\end{eqnarray*}
where $R_\vecn$ is a rotation that turns vector $\uparrow$ to $\vecn$.
Together with the positivity condition $E_\uparrow \geq 0$, this implies decomposition
\begin{eqnarray*}
	E_\uparrow = \id_\hil + \sum_{i=1}^{3} r_i \sigma_i,
\end{eqnarray*}
with the Pauli matrices and a real vector $\vect[r] = (r_1,r_2,r_3)$ satisfying $|\vect[r]| \leq 1$.
Since vector $\uparrow$ is invariant under rotations along $z$-axis,
\begin{eqnarray*}
	E_\uparrow = E_{R_z(\uparrow)} = U_{R_z} E_\uparrow U_{R_z}^\dagger,
\end{eqnarray*}
holds for any rotation $R_z$ along $z$-axis, so that $\sigma_1$ and $\sigma_2$ components of $E_\uparrow$ are eliminated.
Finally we have the decomposition
\begin{eqnarray}
\nonumber	E_\uparrow &=& \id_\hil + r_3 \sigma_3 \\
\label{eq:cov_decompo_1} &=& (1+r_3) \ketbra[\uparrow] + (1-r_3) \ketbra[\downarrow],
\end{eqnarray}
with a real number $r_3 \in [-1,1]$.
For this POVM operator, the probability (density) on state $\rhne$ reduces to
\begin{eqnarray}
\label{eq:prob_cov_1}	\trace[ \rhne E_\uparrow] = (1 -\delta) r_3 \cos \theta +1
\end{eqnarray}
where $\theta$ represents the angle between $z$-axis and $\vecn$.

Now the integration for fidelity $f$ can be calculated to have
\begin{eqnarray*}
	f && (\ensrhn', \{ E_{\vect[m]} \}_{\vect[m] \in S_2}, g) \\
	&&= \int \dif \vecn \dif \vect[m] \trace[ \rhne E_{\vect[m]} ] \frac{1+\vecn \cdot \vect[m]}{2} \\
	&&= \int \dif \vecn \dif \vect[m] \trace[ \rhne U_{R_{\vect[m]}} E_\uparrow U_{R_{\vect[m]}}^\dagger] \frac{1+R^{-1}_{\vect[m]}(\vecn) \cdot \uparrow}{2} \\
	&&= \int \dif \vecn \dif \vect[m] \trace[ \rho_{R^{-1}_{\vect[m]}(\vecn),\delta} E_\uparrow] \frac{1+R^{-1}_{\vect[m]}(\vecn) \cdot \uparrow}{2} \\
	&&= \int \dif \vecn \trace[ \rhne E_\uparrow] \frac{1+\vecn \cdot \uparrow}{2} \\
	&&= \frac{1}{2} + \frac{r_3(1-\delta)}{6}.
\end{eqnarray*}
The maximum value of $f$ is $2/3 - \delta/6$ obtained for $r_3 = 1$.
Finally, we have derived the averaged fidelity
\begin{eqnarray*}
	\fidelity (\enshfe) = \fidelity (\ensrhn') = \frac{2}{3} - \frac{\delta}{6},
\end{eqnarray*}
accomplished by the optimal covariant measurement with POVM elements $\{ 2 \ketbra[\vecn] \}_{\vecn \in S_2}$.

\subsection{Achievable averaged fidelity of $\enshfe$ with its single copy}\label{sec:lowerboundfidelity}
Here we describe our measurement strategy for obtaining the fidelity $f_\sr (\enshfe) = (2 \sqrt{3} + 15)/24 - (2 \sqrt{3}+3)\delta/24$ of $\enshfe$ with its single copy.
Let us first define vectors $\vecn_i$ $(i=0,1,2,3)$ pointing to the summits of tetrahedron by
\begin{eqnarray*}
	\vecn_0 &=& (0,0,1),~	\vecn_1 = \left( \frac{2 \sqrt{2}}{3},0,- \frac{1}{3} \right),\\
	\vecn_2 &=& \left(- \frac{ \sqrt{2}}{3},\sqrt{\frac{2}{3}}, - \frac{1}{3} \right),~	\vecn_3 = \left(- \frac{ \sqrt{2}}{3},- \sqrt{\frac{2}{3}}, - \frac{1}{3} \right).
\end{eqnarray*}
Following unit vectors in $\hil \otimes \hil$ are employed from \cite{Massar1995,GisinPopescu1999} for the measurement strategies:
\begin{eqnarray*}
	\ket[\para^+_i] &:=& \frac{\sqrt{3}}{2} e^{i \phi^+} \ket[\vecn_i] \otimes \ket[\vecn_i] + \frac{1}{2} \ket[\Psi^-], \\
	\ket[\para^-_i] &:=& \frac{\sqrt{3}}{2} e^{i \phi^-} \ket[-\vecn_i] \otimes \ket[-\vecn_i] + \frac{1}{2} \ket[\Psi^-], \\
	\ket[\anpa^+_i] &:=& a \ket[\vecn_i] \otimes \ket[- \vecn_i] - b \sum_{j \neq i} \ket[\vecn_j] \otimes \ket[- \vecn_j],\\
	\ket[\anpa^-_i] &:=& a \ket[- \vecn_i] \otimes \ket[ \vecn_i] - b \sum_{j \neq i} \ket[- \vecn_j] \otimes \ket[ \vecn_j].
\end{eqnarray*}
where $a=(3\sqrt{3}+1)/4\sqrt{2}$, $b=(\sqrt{3} - 1)/4\sqrt{2}$ and $\ket[\Psi^-]$ is the antisymmetric state.
Spin direction $\vecn= (\sin \theta \cos \phi , \sin \theta \cos \phi , \cos \theta)$ is the Bloch vector for state $\ket[\vecn]$ so that $\ket[\vecn] = \cos (\theta /2) \ket[0] + \sin (\theta/2) e^{i \phi} \ket[1]$.
Phases $\phi^\pm$ are chosen so that $\ket[\para^\pm_i]$ ($i=1,2,3,4$) are orthogonal to each other.

In the reminder of this subsection, Hilbert spaces are lined in the order $\hil \otimes \hil \otimes \hil' \otimes \hil'$ on the notation.
The POVM operators $\{ E_i \}_{i \in I_4}$ ($I_4 = \{ 0,1,2,3 \}$) for our strategy are given by
\begin{eqnarray}
\label{eq:povm_anomalous}	E_i := E_{00i} + E_{01i} + E_{10i} + E_{11i} ~(\forall i \in I_4),
\end{eqnarray}
where
\begin{eqnarray*}
	E_{00i} &:=& \ketbra[\para^+_i] \otimes \ketbra[0] \otimes \ketbra[0], \\
	E_{11i} &:=& \ketbra[\para^-_i] \otimes \ketbra[1] \otimes \ketbra[1], \\
	E_{01i} &:=& \ketbra[\anpa^+_i] \otimes \ketbra[0] \otimes \ketbra[1], \\
	E_{10i} &:=& \ketbra[\anpa^-_i] \otimes \ketbra[1] \otimes \ketbra[0].
\end{eqnarray*}
The guessing function $g$ is defined by
\begin{eqnarray*}
	g(i) = \mathbf{g}_i = \vecn_i
\end{eqnarray*}
The fidelity $f_\sr$ for $\enshfe$ with this measurement strategy is
\begin{eqnarray}
\nonumber	f_\sr && (\enshfe, \{ E_i \}_{i \in I_4}, g) \\
\nonumber	=&& \sum_{i=0}^3 \int \dif \vecn \trace[\hfe \otimes \hfe E_i] \frac{1+\vecn \cdot \vecn_i}{2} \\
\nonumber	=&& \sum_{i=0}^3 \int \dif \vecn \frac{\bra[\para^+_i] \rhne \otimes \rhne \ket[\para^+_i] }{4} \frac{1+\vecn \cdot \vecn_i}{2} \\
\nonumber	&&+ \sum_{i=0}^3 \int \dif \vecn \frac{\bra[\anpa^+_i] \rhne \otimes \rhmne \ket[\anpa^+_i] }{4} \frac{1+\vecn \cdot \vecn_i}{2} \\
\nonumber	&&+ \sum_{i=0}^3 \int \dif \vecn \frac{\bra[\para^-_i] \rhmne \otimes \rhmne \ket[\para^-_i] }{4} \frac{1+\vecn \cdot \vecn_i}{2} \\
\nonumber	&&+ \sum_{i=0}^3 \int \dif \vecn \frac{\bra[\anpa^-_i] \rhmne \otimes \rhne \ket[\anpa^-_i] }{4} \frac{1+\vecn \cdot \vecn_i}{2} \\
\nonumber	=&& \int \dif \vecn \bra[\para^+_0] \rhne \otimes \rhne \ket[\para^+_0] (1+\vecn \cdot \vecn_0) \\
\label{eq:integral_fidelity}	&&+ \int \dif \vecn \bra[\anpa^+_0] \rhne \otimes \rhmne \ket[\anpa^+_0] (1+\vecn \cdot \vecn_0),
\end{eqnarray}
where the last equality comes from the symmetry of the measurement.
Under the parameter $\vecn = (\sin \theta \cos \phi, \sin \theta \sin \phi, \cos \theta)$, we have
\begin{eqnarray}
\nonumber	&& \bra[\para^+_0] \rhne \otimes \rhne \ket[\para^+_0] \\
\nonumber	&&= \frac{3(1-\delta)^2 \cos^2 \theta + 6(1-\delta)\cos \theta -(\delta+1)(\delta -3)}{16}, \\
\label{eq:prob_para}	&&\\
\nonumber	&& \bra[\anpa^+_0] \rhne \otimes \rhmne \ket[\anpa^+_0] \\
\nonumber	&&= \frac{3(1-\delta)^2 \cos^2 \theta + 2\sqrt{3} (1-\delta)\cos \theta -(\delta^2 -2\delta -1)}{8},\\
\label{eq:prob_anpa} &&
\end{eqnarray}
which are substituted to the last line of Eq.~(\ref{eq:integral_fidelity}) to yield
\begin{eqnarray*}
	f_\sr (\enshfe, \{ E_i \}_{i \in I_4}, g) = \frac{2 \sqrt{3} + 15}{24} - \frac{2 \sqrt{3}+3}{24} \delta.
\end{eqnarray*}

\section{Mutual information}\label{sec:2}
In this section we calculate mutual information for carriers $\rhn$ and $\hfe$ obtained by the optimal measurements for averaged fidelity.
We derive these values as maximum mutual information accomplished by covariant measurements and then show that the covariant measurement is equivalent to the optimal measurement for averaged fidelity.
Since averaged fidelity for $\enshfe$ with its copy is not obtained, we calculate the corresponding mutual information for $\enshfe$ with its copy from the measurement strategy described in Sec.~\ref{sec:lowerboundfidelity}.

In general, mutual information $H_i(S_2;Y)_{\E_\mu, \{ E_y \}_{y \in Y}}$ ($i=1,2$) between random variables $S_2$ and $Y$ generated by POVM measurement $\{ E_y \}_{y \in Y}$ on ensemble $\E_\mu = \{ \mu_\vecn , \dif \vecn \}_{\vecn \in S_2}$ is given by
\begin{eqnarray*}
&& H(S_2;Y)_{\E_\mu, \{ E_y \}_{y \in Y}} \\
&=& \int \dif \vecn \dif y  \trace[ \mu_\vecn E_y] \log \frac{\trace[ \mu_\vecn E_y]}{\trace[E_y \int \dif \vecn \mu_\vecn]}, \\
&& H_\sr(S_2;Y)_{\E_\mu, \{ E_y \}_{y \in Y}} \\
&=& \int \dif \vecn \dif y \trace[ \mu_\vecn \otimes \mu_\vecn E_y] \log  \frac{ \trace[ \mu_\vecn \otimes \mu_\vecn E_y] }{\trace[E_y \int \dif \vecn \mu_\vecn \otimes \mu_\vecn]}.
\end{eqnarray*}
If the measurements are assumed to be covariant measurement $\{ E_{\vect[m]} \}_{\vect[m] \in S_2}$, $H(S_2;Y)_{\E_\mu, \{ E_y \}_{y \in Y}}$ further simplifies to
\begin{eqnarray}
\nonumber	&&H(S_2;S_2)_{\E_\mu, \{ E_\vecn \}_{\vecn \in S_2}} \\
\nonumber	&=& \int \dif \vecn \dif \vect[m]  \trace[ \mu_\vecn E_{\vect[m]}] \log  \frac{\trace[ \mu_\vecn E_{\vect[m]}]}{\trace[E_{\vect[m]} \int \dif \vecn \mu_\vecn]} \\
\nonumber	&=& \int \dif \vecn \dif \vect[m]  \trace[ \mu_{R^{-1}_{\vect[m]} (\vecn)} E_\uparrow] \log  \frac{\trace[ \mu_{R^{-1}_{\vect[m]}(\vecn)} E_\uparrow]}{\trace[E_\uparrow \int \dif \vecn \mu_{R^{-1}_{\vect[m]} (\vecn)} ]} \\
\label{eq:cov_mut_1}	&=& \int \dif \vecn  \trace[ \mu_\vecn E_\uparrow] \log \frac{\trace[ \mu_\vecn E_\uparrow]}{\trace[E_\uparrow \int \dif \vecn \mu_\vecn ]},
\end{eqnarray}
and $H_\sr(S_2;Y)_{\E_\mu, \{ E_y \}_{y \in Y}}$ to
\begin{eqnarray}
\nonumber	&& H_\sr(S_2;S_2)_{\E_\mu, \{ E_\vecn \}_{\vecn \in S_2}} \\
\label{eq:cov_mut_2}	&=& \int \dif \vecn  \trace[ \mu_\vecn \otimes \mu_\vecn E_\uparrow] \log \frac{\trace[ \mu_\vecn \otimes \mu_\vecn E_\uparrow]}{\trace[E_\uparrow \int \dif \vecn \mu_\vecn \otimes \mu_\vecn]},
\end{eqnarray}
by the same procedure.

\subsection{Mutual information for $\ensrhn$ and $\enshfe$ without copies by optimal covariant measurements}
As we have already noted in Sec.~\ref{sec:stat}, any measurement on $\hfe$ can be simulated by a measurement on $\rhne$, and vice versa.
This relationship is preserved by the restriction to covariant measurements, namely, any covariant measurement on $\hfe$ can be simulated by a covariant measurement on $\rhne$, and vice versa.
In fact, if $U \in \B(\hil)$ is a two dimensional representation of an element from $SU(2)$ acting on $\rhne$, the action of the same element on state $\hfe$ is presented by $U \otimes \ketbra[0] + \sigma_2 U^\ast \sigma_2 \otimes \ketbra[1] \in \B(\hil \otimes \hil')$.
These two representations are interchanged to each other by the statistic morphisms $\ptp_0$ and $\mathcal{J}$ defined by Eqs.~(\ref{eq:def_ptp0}) and (\ref{eq:inverse_statistic}), respectively.
This implies covariant measurements on $\rhne$ and $\hfe$ are also interchanged by these two statistic morphisms.

Thus optimization of the covariant measurement for mutual information of $\enshfe$ can be replaced to that of $\ensrhn'$.
Explicitly, we have
\begin{eqnarray*}
	\max_{\{ E_\vecn \}_\sphere :\mathrm{covariant}} && H(S_2;S_2)_{\enshfe, \{ E_\vecn \}_{\vecn \in S_2}} \\
	&&= \max_{\{ E_\vecn \}_\sphere :\mathrm{covariant}} H(S_2;S_2)_{\ensrhn', \{ E_\vecn \}_{\vecn \in S_2}},
\end{eqnarray*}
and shall consider optimal covariant measurement for $\ensrhn'$.
The analysis includes the optimization of covariant measurement for mutual information of $\ensrhn$ as the special case $\delta = 0$.

\begin{figure}[htbp]
	\includegraphics[width=8.6cm]{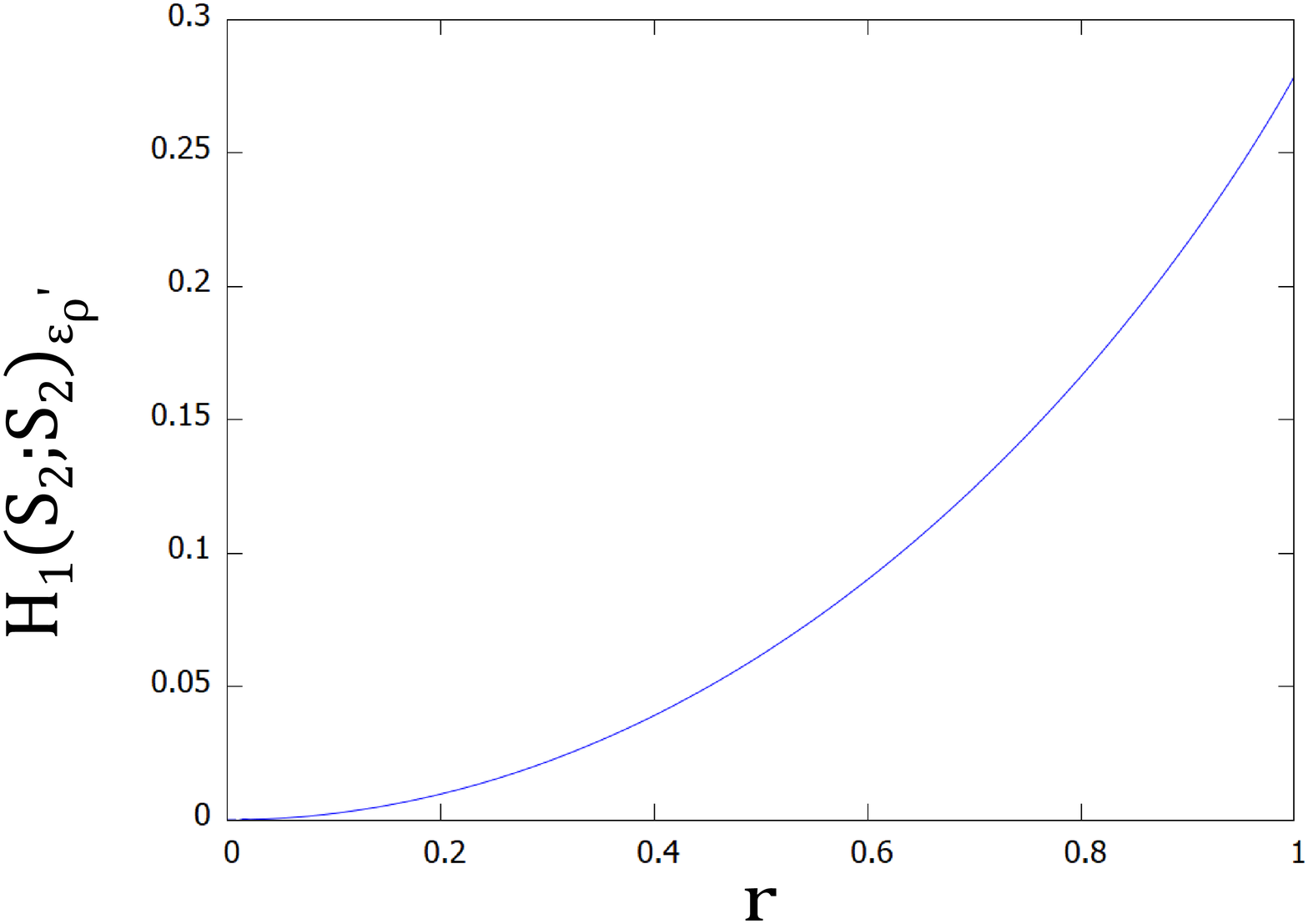}
	\caption{\label{fig:mutinfo1} Mutual information $H(S_2;S_2)_{\ensrhn', \{ E_\vecn \}_{\vecn \in S_2}}$ given by Eq.~(\ref{eq:integral_1}).}
\end{figure}

As already derived, POVM operator $E_\uparrow$ for covariant measurement on $\rhne$ has a decomposition given by Eq.~(\ref{eq:cov_decompo_1}).
The denominator in logarithm of mutual information (\ref{eq:cov_mut_1}) is
\begin{eqnarray*}
	\trace[E_\uparrow \int \dif \vecn \rhne ] = \trace[E_\uparrow \frac{\id_\hil}{2} ] = 1,
\end{eqnarray*}
where we used $\int \dif \vecn \rhne = \id_\hil/2$.
Then the mutual information is
\begin{eqnarray*}
	H(S_2;S_2)_{\E_\mu, \{ E_\vecn \}_{\vecn \in S_2}} = \int \dif \vecn  \trace[ \rhne E_\uparrow] \log \trace[ \rhne E_\uparrow] 
\end{eqnarray*}
This integral can be calculated by substituting Eq.~(\ref{eq:prob_cov_1}) which yields
\begin{eqnarray}
\nonumber	H(S_2;S_2)_{\ensrhn', \{ E_\vecn \}_{\vecn \in S_2}}	&&= \frac{1}{2} \left(\frac{r}{2} + 1 + \frac{1}{2r} \right) \log (1+r) \\
\label{eq:integral_1}	- \frac{1}{2} \left( \frac{r}{2} - 1 + \frac{1}{2r} \right) && \log (1-r) - \frac{1}{2 \ln 2},
\end{eqnarray}
where $r=(1-\delta)r_3$.
Mutual information (\ref{eq:integral_1}) is monotonically increasing according to $|r|$.
Figure~\ref{fig:mutinfo1} presents $H(S_2;S_2)_{\ensrhn', \{ E_\vecn \}_{\vecn \in S_2}}$ as function (\ref{eq:integral_1}) of $r$.
The maximum value of (\ref{eq:integral_1}) is obtained when $r_3 = 1$, and hence $r = 1- \delta$, in which case the measurement is also optimal for averaged fidelity (see Appendix~\ref{sec:fidelity_singletau}).
We finally arrive at the values
\begin{eqnarray*}
	\max_{\{ E_\vecn \}_{\vecn \in S_2}:\mathrm{covariant}} && H(S_2;S_2)_{\enshfe, \{ E_\vecn \}_{\vecn \in S_2}} \\
	= \frac{1}{2} && \left(\frac{1-\delta}{2} + 1 + \frac{1}{2(1-\delta)} \right) \log (2-\delta) \\
	- \frac{1}{2} && \left( \frac{1-\delta}{2} - 1 + \frac{1}{2(1-\delta)} \right) \log \delta - \frac{1}{2 \ln 2},
\end{eqnarray*}
for ensemble $\enshfe$ and
\begin{eqnarray*}
	\max_{\{ E_\vecn \}_{\vecn \in S_2}:\mathrm{covariant}} && H(S_2;S_2)_{\ensrhn, \{ E_\vecn \}_{\vecn \in S_2}} \\
	&&= 1- \frac{1}{2 \ln 2} \approx  0.279,
\end{eqnarray*}
for ensemble $\ensrhn$, by taking $\delta=0$.

\subsection{Mutual information for $\ensrhn$ with its copy by optimal covariant measurement}
According to \cite{Massar2000}, the $+z$ component $E_\uparrow$ of POVM operators $\{ E_\vecn \}_{\vecn \in S_2}$ for covariant measurement on $\rhn \otimes \rhn$ is represented by
\begin{eqnarray*}
	E_\uparrow &=& \id_{\hil} \otimes \id_{\hil} + \alpha (\sigma_3 \otimes \id_\hil + \id_\hil \otimes \sigma_3) \\
	&+& \gamma (2 \sigma_3 \otimes \sigma_3 - \sigma_1 \otimes \sigma_1 - \sigma_2 \otimes \sigma_2)
\end{eqnarray*}
where two real parameters $\alpha$ and $\gamma$ satisfy
\begin{eqnarray*}
	\alpha \leq \frac{\gamma}{2} + 1,~ \alpha \geq - \frac{\gamma}{2} - 1,~\gamma \leq 1,
\end{eqnarray*}
for operator $E_\uparrow$ to be positive.
We substitute this decomposition of $E_\uparrow$ to Eq.~(\ref{eq:cov_mut_2}) to calculate the mutual information of $\ensrhn$ obtained by covariant measurements on the $2$ copies.

The denominator in logarithm of mutual information (\ref{eq:cov_mut_2}) is
\begin{eqnarray*}
	\trace[E_\uparrow \int \dif \vecn \rhn \otimes \rhn] &=& \trace[\int \dif \vecn E_{R^{-1}_\vecn(\uparrow)}  \rho_\uparrow \otimes \rho_\uparrow] \\
	&=& \trace[\rho_\uparrow \otimes \rho_\uparrow] =1,
\end{eqnarray*}
and thus we have
\begin{eqnarray}
\nonumber	H_\sr && (S_2;S_2)_{\ensrhn, \{ E_\vecn \}_{\vecn \in S_2}} \\
\label{eq:bef_integral}	&&= \int \dif \vecn  \trace[ \rhn \otimes \rhn E_\uparrow] \log \trace[ \rhn \otimes \rhn E_\uparrow].
\end{eqnarray}
The probability (density) $\trace[ \rhn \otimes \rhn E_\uparrow]$ is given by
\begin{eqnarray}
\label{eq:prob}	\trace[ \rhn \otimes \rhn E_\uparrow] = \frac{3 \gamma}{4} \cos^2 \theta + \alpha \cos \theta +1 -\frac{\gamma}{4}
\end{eqnarray}
where $\theta$ is the angle between $\uparrow$ and $\vecn$ \cite{Massar2000}.

\begin{figure*}[htbp]
	\includegraphics[width=\textwidth]{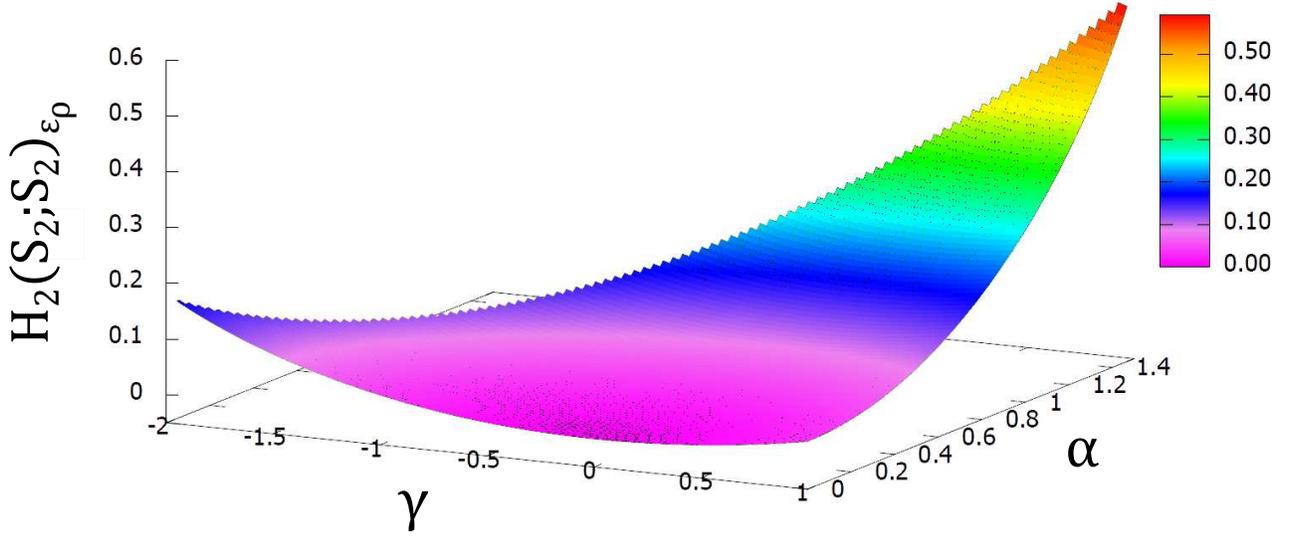}
	\caption{\label{fig:mutinfo2_3d} Mutual information $H_\sr(S_2;S_2)_{\ensrhn, \{ E_\vecn \}_{\vecn \in S_2}}$ given as a function of $\alpha$ and $\gamma$ is plotted in region $\gamma \leq 1$, $\alpha \leq \gamma/2 + 1$, and $\alpha \geq 0$. While the graph at negative $\alpha$ is omitted here, $H_\sr(S_2;S_2)_{\ensrhn, \{ E_\vecn \}_{\vecn \in S_2}}$ is symmetric with respect to the transformation $\alpha \rightarrow - \alpha$.}
\end{figure*}

The analytical solution of the integral (\ref{eq:bef_integral}) needs case divergence based on the values of $\alpha$ and $\gamma$.
Let $D(\alpha,\gamma)$ be the discriminator
\begin{eqnarray*}
	D(\alpha,\gamma) = \alpha^2 + \frac{3}{4} \gamma^2 - 3 \gamma
\end{eqnarray*}
of Eq.~(\ref{eq:prob}) seen as an equation of order $2$ of $\cos \theta$.
Then $H_\sr(S_2;S_2)_{\ensrhn, \{ E_\vecn \}_{\vecn \in S_2}}$ is equal to
\begin{eqnarray*}
	\frac{(1+\alpha)^2}{2 \alpha} \log (1+\alpha) - \frac{(1-\alpha)^2}{2 \alpha} \log (1-\alpha) -1, 
\end{eqnarray*}
when $\gamma = 0$, otherwise to
\begin{eqnarray*}
	h_0(\alpha,\gamma) :=&& \left( \frac{\alpha}{3} - \frac{4 \alpha^3}{27 \gamma^2} + \frac{2 \alpha}{3 \gamma} +1 \right) \log \left( \frac{\gamma}{2} +1 +\alpha \right) \\
	&&- \left( \frac{\alpha}{3} - \frac{4 \alpha^3}{27 \gamma^2} + \frac{2 \alpha}{3 \gamma} -1 \right) \log \left( \frac{\gamma}{2} +1 -\alpha \right) \\
	&&+ \frac{1}{\ln 2} \left( \frac{\gamma}{3} + \frac{4 \alpha^2}{9 \gamma} - \frac{8}{3} \right),
\end{eqnarray*}
when $D(\alpha,\gamma)=0$ and to
\begin{eqnarray*}
	&& h_0(\alpha,\gamma) + \frac{8(-D(\alpha,\gamma))^{\frac{3}{2}}}{27 \gamma^2 \ln 2} \\
	&& \times \left(  \arctan \frac{\alpha + \frac{3 \gamma}{2}}{\sqrt{-D(\alpha,\gamma)}} - \arctan \frac{\alpha - \frac{3 \gamma}{2}}{\sqrt{-D(\alpha,\gamma)}} \right)
\end{eqnarray*}
when $D(\alpha,\gamma) < 0$ and to
\begin{eqnarray*}
	h_0(\alpha,\gamma) + \frac{4D(\alpha,\gamma)^{\frac{3}{2}}}{27 \gamma^2 } \log \frac{1- \gamma + \sqrt{D(\alpha,\gamma)}}{1- \gamma - \sqrt{D(\alpha,\gamma)}} ,
\end{eqnarray*}
when $D(\alpha, \gamma) >0$.
Note that when $D(\alpha,\gamma) >0$, the solution can be written in a different form
\begin{widetext}
\begin{eqnarray*}
	\left( \frac{\alpha}{3} - \frac{4 \alpha^3}{27 \gamma^2} + \frac{2 \alpha}{3 \gamma} +1 - \frac{4D(\alpha,\gamma)^{\frac{3}{2}}}{27 \gamma^2 } \right) &&  \log  \left( \frac{\gamma}{2} +1 +\alpha \right) \\
	+ \left( - \frac{\alpha}{3} + \frac{4 \alpha^3}{27 \gamma^2} - \frac{2 \alpha}{3 \gamma} +1 - \frac{4D(\alpha,\gamma)^{\frac{3}{2}}}{27 \gamma^2 } \right) && \log  \left( \frac{\gamma}{2} +1 -\alpha \right) + \frac{1}{\ln 2} \left( \frac{\gamma}{3} + \frac{4 \alpha^2}{9 \gamma} - \frac{8}{3} \right) \\
	+ \frac{4D(\alpha,\gamma)^{\frac{3}{2}}}{27 \gamma^2 } \log && (1- \gamma + \sqrt{D(\alpha,\gamma)})^2
\end{eqnarray*}
\end{widetext}
from which convergences along lines $\alpha = \gamma/2 + 1$ and $\alpha = - \gamma/2 - 1$ are easier to be seen.
Figure~\ref{fig:mutinfo2_3d} presents $H_\sr(S_2;S_2)_{\ensrhn, \{ E_\vecn \}_{\vecn \in S_2}}$ as function of $\alpha$ and $\gamma$.
The maximum of mutual information is obtained for
\begin{eqnarray*}
	\alpha = \pm \frac{3}{2}, ~ \gamma = 1,
\end{eqnarray*}
at which the covariant measurement is also optimal for averaged fidelity on two copies \cite{Massar2000}.
We finally obtain
\begin{eqnarray}
\nonumber	\max_{\{ E_\vecn \}_{\vecn \in S_2}:\mathrm{covariant}}&& H_\sr(S_2;S_2)_{\ensrhn, \{ E_\vecn \}_{\vecn \in S_2}} \\
\label{eq:max_cov_mut_rhn2}	&&= \log 3 - \frac{2}{3 \ln 2} \approx  0.623.
\end{eqnarray}

\subsection{Mutual information for $\enshfe$ with its single copy}
The measurement strategy we employed for deriving averaged fidelity of $\enshfe$ exceeding that of $\ensrhn$ with their copies is not shown to be the optimal covariant measurement.
Here we calculate mutual information of $\enshfe$ obtained by the same measurement strategy. 

The random variable of observer's register is $I_4:=\{ 0,1,2,3 \}$, and the POVM operators $\{ E_i \}_{i \in I_4}$ is presented in Eq.~(\ref{eq:povm_anomalous}).
Mutual information is given by
\begin{eqnarray}
\nonumber	&& H_\sr (S_2;I_4)_{\enshfe, \{ E_i \}_{i \in I_4}} \\
\nonumber	&=& \sum_{i=0}^3 \int \dif \vecn \trace[\hfe \otimes \hfe E_i] \log \frac{\trace[\hfe \otimes \hfe E_i]}{ \trace[E_i \int \dif \vecn \hfe \otimes \hfe ] } \\
\nonumber	&=& 4 \int \dif \vecn \trace[\hfe \otimes \hfe E_0] \log \frac{\trace[\hfe \otimes \hfe E_0]}{ \trace[E_0 \int \dif \vecn \hfe \otimes \hfe ] } \\
\label{eq:integral} &&
\end{eqnarray}
where the second equality follows from symmetry of the POVM elements.
The denominator in logarithm is
\begin{eqnarray*}
	\trace[E_0 \int \dif && \vecn \hfe \otimes \hfe ] \\
	= \frac{1}{2} && \bra[\para^+_0] \int \dif \vecn \rhne \otimes \rhne \ket[\para^+_0] \\
	&&+ \frac{1}{2} \bra[\anpa^+_0] \int \dif \vecn \rhne \otimes \rho_{- \vecn,\delta} \ket[\anpa^+_0] \\
	= \frac{1}{4} &&.
\end{eqnarray*}
The measurement probability (density) is
\begin{eqnarray*}
	\trace[ \hfe \otimes \hfe && E_0] \\
	= && \frac{1}{2} \bra[\para^+_0]  \rhne \otimes \rhne \ket[\para^+_0] \\
	&&+ \frac{1}{2} \bra[\anpa^+_0] \rhne \otimes \rho_{- \vecn,\delta} \ket[\anpa^+_0] \\
	= && \frac{9(1-\delta)^2 \cos^2 \theta }{32} + \frac{(6+4\sqrt{3})(1-\delta)\cos \theta }{32} \\
	&& + \frac{ \left\{8 - 3(1-\delta)^2 \right\} }{32},
\end{eqnarray*}
where we substitute Eqs.~(\ref{eq:prob_para}) and (\ref{eq:prob_anpa}).
Analytical solution of the integral (\ref{eq:integral}) is given by
\begin{widetext}
\begin{eqnarray*}
	&& \left\{ \frac{3 + 2\sqrt{3}}{24} q \left( 1+ \frac{87-12\sqrt{3}}{81q^2} \right)+\frac{1}{2} \right\} \log \left(  \frac{3}{4}q^2 + \frac{3+2\sqrt{3}}{4}q +1 \right) \\
	&&- \left\{ \frac{3 + 2\sqrt{3}}{24} q \left( 1+ \frac{87-12\sqrt{3}}{81q^2} \right)-\frac{1}{2} \right\} \log \left(  \frac{3}{4}q^2 -\frac{3+2\sqrt{3}}{4}q +1 \right) + \frac{1}{4 \ln 2} \left( q^2 + \frac{4\sqrt{3} -41}{9} \right) \\
	&&+ \frac{(51-12\sqrt{3}-27q^2)^{\frac{3}{2}}}{972 q \ln 2} \left( \arctan  \frac{3+2\sqrt{3}+9q}{\sqrt{51-12\sqrt{3} -27q^2}} - \arctan \frac{3+2\sqrt{3}-9q}{\sqrt{51-12\sqrt{3} -27q^2}} \right),
\end{eqnarray*}
\end{widetext}
where $q=1 -\delta$.
When $\delta=0$ the value of $H_\sr (S_2,I_4)_{\enshfe, \{ E_i \}_{i \in I_4}}$ reduces to
\begin{widetext}
\begin{eqnarray*}
	\frac{117 + 25 \sqrt{3}}{162} \log \frac{5+\sqrt{3}}{2} &&- \frac{-45 +25 \sqrt{3}}{162} \log \frac{2-\sqrt{3}}{2} + \frac{\sqrt{3} - 8}{9 \ln 2}  \\
	+&& \frac{2(6 -3\sqrt{3})^{\frac{3}{2}}}{243 \ln 2} \left( \arctan \frac{6+\sqrt{3}}{\sqrt{6-3\sqrt{3}}} - \arctan \frac{-3+\sqrt{3}}{\sqrt{6-3\sqrt{3}}} \right) \approx 0.718,
\end{eqnarray*}
\end{widetext}
which is greater than mutual information (\ref{eq:max_cov_mut_rhn2}) of $\ensrhn$ obtained by the optimal covariant measurement on two copies.

\section{Existence of statistical transformations between probabilistic carriers}\label{sec:classical}
Let $\E_\tau = \{ \tau_x \in \B(\hil_1), p_x \}_{x \in  X}$ and $\E_\rho = \{ \rho_x \in \B(\hil_2), p_x \}_{x \in  X}$ be probabilistic carriers.
For the carriers to be probabilistic means that the states have decompositions
\begin{eqnarray}
	\tau_x = \sum_i t(i|x) \ketbra[i]~(\forall x \in X),\\
	\rho_x = \sum_j r(j|x) \ketbra[j]~(\forall x \in X),
\end{eqnarray}
for some bases $\{ \ket[i] \}_i$ of $\hil_1$ and $\{ \ket[j] \}_j$ of $\hil_2$, where $t(i|x)$ and $r(j|x)$ are conditional probabilities.
We show that condition (\ref{eq:cond3}) of sNQI implies the existence of statistical transformation $L:\B(\hil_2) \rightarrow \B(\hil_1)$ such that $\trace[E \tau_x] = \trace[L(E) \rho_x]$ for any $x \in X$ and any POVM operator $E$ for probabilistic carriers.

We first show that condition (\ref{eq:cond3}) implies
\begin{description}
	\item[P] for any set of POVM elements $\{ E_y \in \B(\hil_1) \}_{y \in Y}$ there is a set of POVM elements $\{ E'_y \in \B(\hil_2) \}_{y \in Y}$ such that $\trace [ E_y \tau_x ] = \trace[ E'_y \rho_x ]$ holds for any $y \in Y$ and $x \in X$.
\end{description}
Let us write $\E_\tau \preceq \E_\rho$ when $\mathbf{P}$ holds for general ensembles $\E_\tau$ and $\E_\rho$.
Suppose that $\E_\tau \preceq \E_\rho$ does not hold.
Define a SC measure $\ssm$ on probabilistic carriers by
\begin{eqnarray}
	\ssm(\E_\mu) = \left\{ \begin{array}{lc}
	1 & (\E_\tau \preceq \E_\mu)\\
	0 & (\text{Otherwise}).
	\end{array}\right.
\end{eqnarray}
This is a valid SC measure: transitivity of relation $\preceq$ implies that $\ssm$ is well-defined, and it is a function of probabilities obtained by single POVM measurement since the carriers are probabilistic.
Then we have $\ssm(\E_\tau) =1$, $\ssm(\E_\rho) = 0$, and thus $\ssm(\E_\tau) > \ssm(\E_\rho)$.
By contraposition, condition (\ref{eq:cond3}) implies $\E_\tau \preceq \E_\rho$.

Next, let us take $\{ \ketbra[i]  \in \B(\hil_1) \}_i$ as a POVM operator on $\B(\hil_1)$.
Condition $\E_\tau \preceq \E_\rho$ implies the existence of POVM operator $E'_i = \sum_j e'(j,i) \ketbra[j]$ such that $\trace[ \ketbra[i] \tau_x ] = \trace[ E'_i \rho_x ]$ for all $x \in X$.
Now we can define the statistical transformation $L$ satisfying $\trace[E \tau_x] = \trace[L(E) \rho_x]$ by $L(\ketbra[i]) = E'_i$ and the linearity.
This definition is possible since any POVM operator $E$ on probabilistic carriers has the linear decomposition $E = \sum_i e(i) \ketbra[i]$.

\end{document}